\documentclass[aps,prb,showpacs,]{revtex4}
\usepackage{epsfig} 
\usepackage{graphicx}
\begin{document}
\title{Monte Carlo study of the critical temperature for the planar rotator
model with nonmagnetic impurities}
\author{S.A.Leonel}
\author{ Pablo Zimmermann Coura}
\affiliation{Departamento de F\'{\i}sica ICE, Universidade Federal de
 Juiz de Fora, Juiz de Fora, CEP 36036-330,Minas Gerais, Brazil}
\author{ A.R.Pereira}
\author{L.A.S.M\'ol}
\affiliation{Departamento de F\'{\i}sica,Universidade Federal de Vi\c{c}osa,36571-000,
Vi\c{c}osa, Minas Gerais, Brazil}
\author{B.V.Costa}
\affiliation{Departamento de F\'{\i}sica, ICEX, Universidade Federal de Minas Gerais,
Caixa Postal 702, CEP 30123-970, Belo Horizonte, MG, Brazil}

\begin{abstract}
We performed Monte Carlo simulations to calculate the
Berezinskii-Kosterlitz-Thouless (BKT) temperature $T_{BKT}$ for the
two-dimensional planar rotator model in the presence  of nonmagnetic impurity
concentration $(\rho)$. As expected, our calculation shows that the
BKT temperature  decreases as the spin vacancies increase. There is a
critical dilution $\rho_c \approx 0.3$ at which $T_{BKT} =0$.
The effective
interaction between a  vortex-antivortex pair and a static nonmagnetic
impurity is studied analytically.  A simple phenomenological argument based on
the pair-impurity interaction is  proposed to justify the simulations.
\end{abstract}
\maketitle 

%
%
%------------------------Introduction------------------------------------
%
\section{Introduction}
The  planar rotator (PR) model in two dimensions
is a prototype for several physical systems as for example
high temperature superconductors and  granular superconductors.
The PR model
supports topological excitations and although there is no long range order
at any finite temperature, it undergoes a BKT
phase transition driven by the unbinding of vortex-antivortex pairs.
In short the BKT picture of the phase transition is as follows. At low
temperature spin waves are the relevant excitations of the system.
Spin-spin correlation functions fall off slowly with distance, free
vortices do not exist but pairs strongly binded. Vortices pairs can
not disorder the system significantly since they affect only close
spins. As the temperature is rised, the distance between vortex-ativortex
pairs grows until $T_{BKT}$. Then free vortices exist, the system is disordered
and the spin-spin correlation function falls off exponentially.
The hamiltonian describing the model is
\begin{equation}
H = - \sum_{<i,j>} J_{i,j}\vec S_i . \vec S_j,
\label{1}
\end{equation}
where $i$ and $j$ enumerate sites in a square lattice, $J_{i,j}$ is an exchange
coupling and
$\vec S_i = \{ S_i^x,S_i^y \} =
\left| S \right| \{ \cos \theta_i,\sin \theta_i \} $
is a two dimensional spin vector.
Of course,  Hamiltonian(\ref{1}) describes an ideal
system, in which each site of a regular  square lattice is occupied by a
spin vector $\vec S$. However, impurity and/or defects are  present in any
material sample. In fact, the effect of impurities  on superconductors has
been of theoretical and experimental interest in its own  right for a long
time. Particularly, the interaction of topological excitations  with spatial
inhomogeneities is of considerable importance from both
theoretical and applied points of view. For example, solitons near a
nonmagnetic  impurity in 2D antiferromagnets cause observable effects in EPR
experiments\cite{KSC,ZKK}. In this scenario it would be important to study the
effects of the presence of nonmagnetic sites diluted in magnetic materials.
In a recent work, M\'ol, Pereira
and Pires\cite{MPP}, have studied the interaction between a static spin
vacancy and a planar vortex and they have shown that the effective potential
experienced between the two defects is repulsive. It indicates that
the presence of spinless atoms on the magnetic plane may affect the BKT
critical temperature.
The main goal in this paper is to consider the effect of magnetic dilution to
the BKT temperature by using numerical and analytical methods.
To take into account the presence of nonmagnetic impurities in our model
(Eq \ref{1}) we can replace  some spin vector $\vec S$ by a $\vec S=0$ creating
a vacancy at that lattice site.
First we
consider that the spin  vacancies are randomly distributed on the sites of the
lattice. The case in  which the spin vacancies are grouped into a cluster will
also be analyzed in  order to compare with the random case. \\

The paper is organized as follows: in section $II$, we describe the model and
the Monte Carlo (MC) method. In section $III$ we present the MC results.
In section $IV$, the continuum theory is used to study the vortex-pair-impurity
interaction and a simple heuristic argument to justify the MC results is
presented and section $V$ contains a summary and final comments.
%
%----------------------Backgroud------------------------------------
%
\section{Background}
We consider in this work a quenched
site diluted $PR$ model. In order to introduce
dilution we define a variable $\sigma_i$ with the following properties: It is
$1$ if site $i$ is magnetic and $0$ otherwise. To acomodate this changes
we have to modify Eq \ref{1} as
\begin{equation}
H = - J\sum_{<i,j>} \sigma_i \vec S_i . \sigma_j \vec S_j =
- J\sum_{<i,j>} \sigma_i \sigma_j \cos \left( \theta_i - \theta_j  \right).
\label{1.1}
\end{equation}
The precise determination of the BKT temperature is a
difficult task due to absence of sharp peaks in the thermodynamic quantities.
One way to extract $T_{BKT}$ was suggested by Weber and
Minnhagen\cite{MIN1,MIN2} by calculating the helicity modulus defined as
\begin{equation}
\Upsilon = \frac{\partial^2F}{\partial \Delta^2}
\label{gamma1}
\end{equation}
where $F$ is the free energy and
$\Delta$ is a small twist across the system in one direction.
Using Eq \ref{1.1} we get
\begin{equation}
\label{2}
\Upsilon=- \frac{1}{2(N-n)}<H_{xy}> -
 \frac{1}{k_{b}T(N-n)}\left<\left[\sum_{i,j}
 \sigma_i \sigma_j sin(\theta_{i}-\theta_{j})\hat
e_{i,j}.\hat x\right]\right>^{2},
\end{equation}
where $N$ is the volume of the system, $n$ is the number of non-magnetic
sites, $\hat e_{i,j}$ is the vector pointing from site $j$ to site $i$ and $\hat
x$  is a unit vector pointing along the x-direction.
The Kosterlitz
renormalization-group equations\cite{KOS} lead to the  prediction that
$\Upsilon$ jumps from the value $\left(\frac{2}{\pi}\right)T_{c}$ to  zero at
the critical temperature,
\begin{equation}
\label{3}
\lim_{T\rightarrow T_{c}}\frac{\Upsilon}{k_{b}T}=\frac{2}{\pi}.
\end{equation}
To calculate the quantity $\Upsilon$ we use a Monte Carlo (MC) approach using
a standard
Metropolis algorithm with periodic boundary
conditions\cite{MET}.  In order to reach the thermodynamical equilibrium we
performed long runs of size  $100\times L\times L$, where $L$ is the linear
size of the lattice.
The temperature was varied in steps of size $\Delta T=0.1K$.
Each point in our
simulations is the result of the average  over $2\times 10^{5}$ independent
configurations. In  the figures showing the results of our simulations, when
not indicated, the error bars are  smaller than the symbols.

Figure 1 shows the results from MC simulations of $\Upsilon$ for lattices
with $5\%$ of impurities and sizes $L=30, 60$ and $80$. The
straight
line represents $\left(\frac{2}{\pi}\right)T$. The crossing point
between this line and $\Upsilon$ gives an estimate of
the $BKT$ temperature. Of course,
this estimate becomes more accurate as the lattice size
increases. However, as we can see in figure 1, the lattice of size $L=60$
gives already a good result adequated for our purposes. From now on we use
the following, $k_{b}=1$ and  the symbol $T_{BKT}$ is used for $T_{c}(\rho
=0)$, i.e, $T_{BKT}=T_{c}(\rho =0)$.
%
%-------------------------------figure 1 --------------------------------
%
\begin{figure}[th]
\begin{tabular}
[c]{clr}%
\mbox{\hspace{0.5cm}} & \resizebox{!}{6cm}{\includegraphics{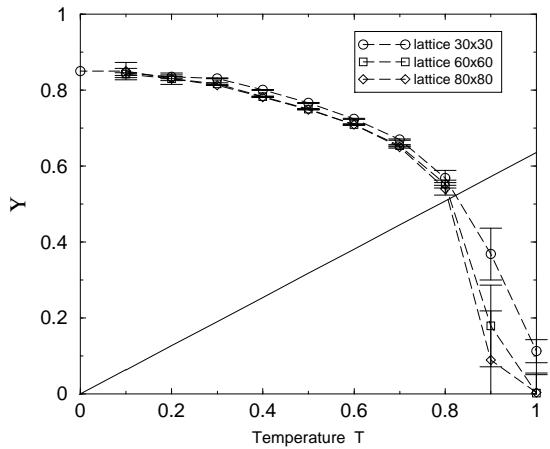}} 
\end{tabular}
\caption{Helicity modulus $\Upsilon$ as a function of temperature for
lattices with sizes 30x30, 60x60,80x80 and with $5\%$ of nonmagnetic
impurities randomly distributed. The solid line is
the curve $\left(\frac{2}{\pi}\right)T$ and the dashed lines are only guides to
the eyes.}
\label{fig1}
\end{figure}
%
%-------------------------------end figure 1--------------------------------
%
%
%-------------------------------Monte Carlo --------------------------------
%
\section{Monte Carlo Results}

\begin{figure}[th]
\begin{tabular}
[c]{clr}%
\mbox{\hspace{0.5cm}} & \resizebox{!}{7cm}{\includegraphics{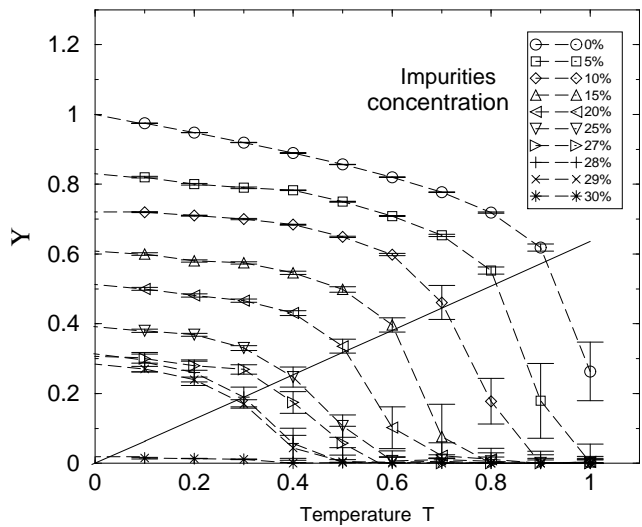}} 
\end{tabular}
\caption{Helicity modulus $\Upsilon$ as a function of temperature for
 lattices size 60x60 with $0\%, 5\%, 10\%, 15\%, 20\%, 25\%, 27\%, 28\%, 29\%$
 and $30\%$ of nonmagnetic impurities randomly distributed. The solid line
 is the line $\left(\frac{2}{\pi}\right)T$ and dashed curves are guides to the eye.}
\label{fig2}
\end{figure}

In this section we present the results obtained by MC simulations.
First, we distribute the
nonmagnetic impurities at random in the lattice sites. Figure \ref{fig2}
contains the helicity modulus as a function of the temperature considering
several values of the impurity concentration $(\rho)$.
It is also shown the straight
line representing the function $\left(\frac{2}{\pi}\right)T$. As noticed
before the intersection of this line with the value of each $\Upsilon$,
gives $T_{c}$ for the corresponding impurity concentration.
%
%-------------------------------figure 2 --------------------------------
%
%
%-------------------------------end figure 2--------------------------------
%
We observe that $T_c(\rho)$
decreases with increasing $\rho$.
Since the helicity modulus is a measurement of the phase correlations of the
system\cite{MIN2},
it is not surprising that these correlations are strongly affected by the 
dilution.
It can be understood as follow: if we remove a spin from the lattice, the
nearest neighbors of that spin will have coordination number of three,
one less than in the bulk. The spins in the boundary
have larger fluctuations than  the spins in the bulk lowering the
spin correlations. We should expect that the fluctuation becomes
appreciable disordering the system for large enough nonmagnetic concentrations
up to a critical value where the $BKT$ temperature goes down to zero. 
In figure 3 we show the $BKT$ temperature as a function of the 
nonmagnetic impurity concentration. Note the abrupt
fall of the critical temperature for $\rho_{c} \approx 0.3$.
%
%-------------------------------figure 3 --------------------------------
%
\begin{figure}[th]
\begin{tabular}
[c]{clr}%
\mbox{\hspace{0.5cm}} & \resizebox{!}{5cm}{\includegraphics{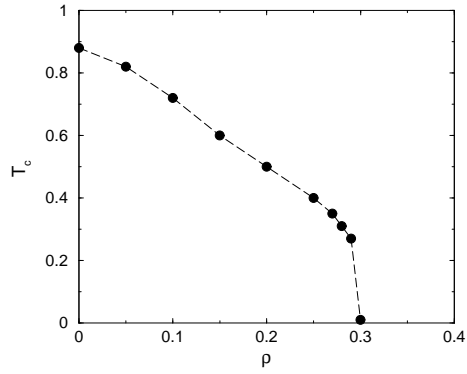}} 
\end{tabular}
\caption{The BKT transition temperature behavior as a function of
 nonmagnetic impurity concentration, based on the MC simulations
 results showed in figure 2. The dashed curves are guides to the eye.}
\label{fig3}
\end{figure}
%
%-------------------------------end figure 3 --------------------------------
%
We also performed MC simulations for the case in which the nonmagnetic
impurities are clustered for $\rho=0.2$ and $0.3$ (see figure \ref{fig4}).
Note that in this case, the critical temperature practically does not
depend on the  impurity concentration $(T_{c}(0.2) \cong T_{c}(0.3))$.
In fact this is an expected result. Since the nonmagnetic cluster
is confined in a region of size $\rho \times L^2$ and the boundary grows as
$\rho \times L$, meaning that spins are still strongly correlated driving the
$BKT$ transition even for large values of $\rho$.
A comparison between the two cases is shown in figure \ref{fig4}.Note
the considerable difference between them. Due to the short range of  the spin
interactions, only the spins near the boundary of the cluster will become 
influenced by the vacancies and hence the correlations of the rest of
the system  will have a behavior almost independent of the vacancies. It must
not affect considerably  the vortices that are formed far way from the cluster
and the phase transition occurs normally.  
%
%-------------------------------figure 4 --------------------------------
%
\begin{figure}
\begin{tabular}
 [c]{clr}%
\mbox{\hspace{0.5cm}} & \resizebox{!}{7cm}{\includegraphics{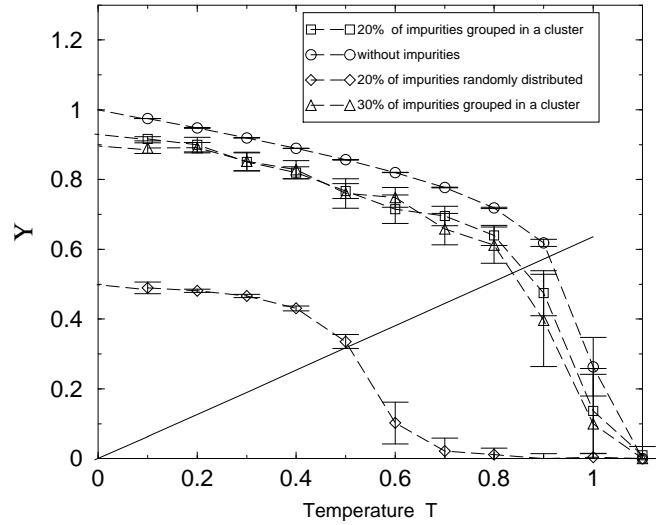}} 
\end{tabular}
\caption{Helicity modulus $\Upsilon$ as a function of temperature,
 for lattices with $20\%$ and $30\%$ of nonmagnetic impurities
 grouped in a cluster, compared with the helicity modulus
 results for lattice with $20\%$ of nonmagnetic impurities
 randomly distributed and lattice without impurities. The
 solid line is the line $\left(\frac{2}{\pi}\right)T$ and
 dashed curves are guides to the eye.}
\label{fig4}
\end{figure}
%
%-------------------------------end figure 4 --------------------------------
%
\section{Vortex-Antivortex-Impurity interaction}
\begin{figure}
\begin{tabular}
[c]{clr}%
\mbox{\hspace{0.5cm}} & \resizebox{!}{5cm}{\includegraphics{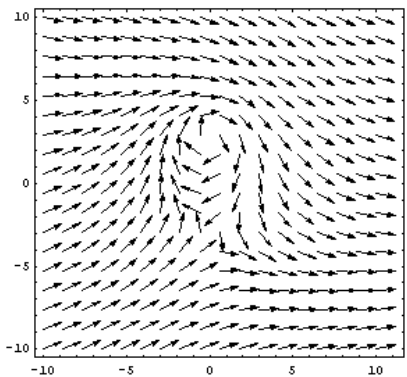}} 
\end{tabular}
\caption{A vortex pair configuration with "center of mass" located
at the origin and size $R=6a$ and a spin vacancy located at (1,3).
The pair configuration is deformed for large distance if vortex 
centers are near the impurity, increasing considerably the system energy.}
\label{fig5}
\end{figure}

In this section we discuss the effect of nonmagnetic sites on the 
vortex-antivortex  structure.
The interaction between the topological excitation and a single nonmagnetic 
impurity below the critical temperature may help us to understand in more 
detail the phase transition mechanism.
In the continuum limit, Hamiltonian (\ref{1}) can be
written as
\begin{equation}
H_{c}=(1/2)J\int (\vec\nabla\theta)^{2}d^{2}x.
\end{equation}
Following reference \cite{MPP}, to take into account the absence of one spin in 
the lattice site  we modify $H_{c}$ as 
\begin{equation}
\label{4}
H_{I}=(1/2)J\int (\vec\nabla\theta)^{2}V(\vec r)d^{2}x  ~~~,
\end{equation}
where $V(\vec r)$ is a localized potential given
by: $V(\vec r)=1$ if $|\vec r-\vec r_{o}|\geq a$, and $V(\vec r)=0$ if $|\vec
r-\vec r_{o}|< a$. Here, the nonmagnetic site is  placed at $\vec r_{o}$ and 
$a$ stands for the 
lattice constant.  This lack of magnetic interaction  inside the circle of
radius $a$, means that a spin located at $\vec r_{o}$ was removed from  the
lattice. The equation of motion obtained from (\ref{4}) is 
\begin{equation}
\label{5}
V(\vec r)\nabla^{2}\theta=-\vec\nabla V(\vec r).\vec\nabla\theta.
\end{equation}
In polar coordinates, the vectors $\vec r$ and $\vec r_{o}$ are written as 
$(r,\phi)$ and
$(r_{o},\phi_{o})$respectively. Then, the gradient of the potential is
\begin{equation}
\label{6}
\vec\nabla V(\vec r)=a[\hat rcos(\alpha-|\phi-\phi_{o}|)+\hat\phi
 sin(\alpha-|\phi-\phi_{o}|)]\delta(\vec r-\vec r_{o}-\vec a),
\end{equation}
where $\delta$ is the Dirac delta function and $\alpha$ is the angle that
the vector $\vec a$, with origin at the point $\vec r_{o}$ and end at a
point on the circumference of the potential ($|\vec a|=a$), makes with the
vector $\vec r_{o}$. In the limit $a\rightarrow 0$, we write
\begin{equation}
\label{7}
\vec\nabla V(\vec r)\approx a[\hat rcos(\alpha)+\hat\phi
sin(\alpha)]\delta(\vec r-\vec r_{o}),
\end{equation}
where $cos(\alpha)$ and $sin(\alpha)$ are anisotropic coupling constants.
A vortex-antivortex pair solution with "center of mass" at the origin is 
given
by $\theta_{2\nu}=arctan[(y-P)/x]-arctan[(y+P)/x]$, where $R=2P$ is the
distance  between the vortex centers. The energy of a pair is
$E_{2\nu}=\pi^{2}J+2\pi J ln(R/a)$.  Note that the energy $E_{2\nu}$ increases
with increasing R implying an attractive  force between vortices of
opposite sign.
Suppposing $\theta_1$ is the deformation introduced in the vortex-antivortex 
structure by the absence of a spin at $\vec r_{o}$ we can write
$\theta=\theta_{2\nu}+\theta_{1}$. 
Rewriting $\theta_{2\nu}$ in polar coordinates and substituting 
$\theta=\theta_{2\nu}+\theta_{1}$ into Eq.(\ref{5}), we obtain
\begin{equation}
\label{8}
\theta_{1}=-\frac{\beta}{\pi}ln\left(\frac{|\vec r - \vec r_{0}|}{a}\right),
\end{equation}
where
\begin{equation}
\label{9}
\beta=\frac{Pa[rr_{0}^{2}cos(2\phi_{0}-\phi)+
rP^{2}cos\phi-r^{3}_{0}cos\phi_{0}-
P^{2}r_{0}cos\phi_{0}]}{|\vec r-\vec r_{0}|(P^{4}+
r_{0}^{4}+2P^{2}r_{0}^{2}cos(2\phi_{0}))}.
\end{equation}
The configuration of this deformed vortex-antivortex pair  is
shown in figure \ref{fig5}.  As the vortex (or antivortex) center approaches
the nonmagnetic impurity, the pair structure becomes more and more
deformed, indicating that there is a  repulsive interaction
potential between each vortex core and the spin vacancy. This
phaenomenum is in
agreement with the results of Ref.\cite{MPP}, where the calculations
took in consideration a single vortex.
%
%-------------------------------figure 5 --------------------------------
%
%
%-------------------------------end figure 5  --------------------------------
%
In order to understand how is
the effective interaction potential between the two defects, we
substitute $\theta=\theta_{2\nu}+\theta_{1}$ into Eq.(\ref{4})
to calculate the energy of the pair-impurity system $E_{PI}$. 
Unfortunately, the integral in Eq.(\ref{4}) can not be done 
analytically for a general impurity position, but in the
special case the spin
vacancy is located at the center of mass
we can solve
it exactly. Using the dominant terms, the effective potential is given by
\begin{equation}
\label{10}
V_{eff}\approx \frac{a^{2}J}{2\pi
P^{2}}\left[\frac{1}{3}ln^{3}\left(\frac{d}{a}\right)+
ln\left(\frac{d}{a}\right)+\frac{4\pi d}{a}\right],
\end{equation}
where $V_{eff}=E_{PI}-E_{2\nu}$, and $d$ is the lattice size. 
Since $P$ is the distance between the vortex (or antivortex)
center and the spin vacancy, this expression is very alike
with the effective potential obtained in Ref.\cite{MPP}, 
between a single vortex and a nonmagnetic impurity. Note
that the effective interaction potential increases with 
decreasing $R=2P$, implying a repulsive force between 
vortices and impurities. In fact, the spin vacancy force
obtained from Eq.(\ref{10}) acts as a "repulsive force"
weakening the coupling strength between the bound vortices,
and becomes stronger as $P$ decreases. Here, the nonmagnetic 
impurity must repel simultaneously the two vortices in a pair, 
affecting the spin field for large distances (see fig. 5).
For a lattice of size $d$, the effective potential (\ref{10})
is a minimum only if $P\rightarrow d/2(R\rightarrow d)$, showing
the tendency of a complete separation of the vortices in a pair
due to the presence of the vacancy. We conclude that static
spin vacancies repel vortices, independently if they are free
or bound into pairs.
Based on the above results, we propose
a phenomenological model to explain the behavior of the $BKT$
temperature as a function of the impurity concentration. 
As discussed above a nonmagnetic site can induce a
a repulsive potential between a pair vortex-antivortex in such a way we can have 
the two scenario. If the nonmagnetic
impurity is in between the pair vortex-antivortex the 
effective repulsive potential created tends to unbind the pair.
On the other hand, if the impurity is not in
between the pair the force in the
nearest vortex will be stronger than in the other and the
tendency is to increase the vortex-antivortex attraction
leading to the annihilation of the pair. Then, impurities may induce
either vortex-antivortex unbinding process or pair annihilation. 
In a system  containing a random distribution
of impurities one can expect a lower density of vortices at any
temperature than in a pure system due to the annihilation 
of pairs. Beside that, the unbinding of vortices-antivortices
should occurs at lower temperature inducing the $BKT$ transition.

Hence, we may expect a critical nonmagnetic impurity
concentration in which vortex pairs 
are not more formed and  the
BKT critical temperature goes to zero. 
The situation is different for the case in which the nonmagnetic
impurities are clustered. In this case, vortex pairs
will be excited far way from the cluster in order to minimize their
energies and the cluster would have only a small influence on the
vortex-antivortex unbinding. The critical temperature should not
be much affected. The results presented in figure \ref{fig4} confirms this 
conjecture. 
%
%-------------------------------Conclusion --------------------------------
%
\section{Summary}
We have performed Monte Carlo simulation for the diluted planar rotator model
in a square lattice. We have found that the $BKT$
temperature decreases with increasing impurity concentration and that
there is a critical impurity concentration $\rho_c \cong 0.3$  at which the 
transition temperature goes to zero.
The interaction
between a vortex-pair and a static spin vacancy was studied 
in the continuum approximation. By considering the decoupling
of vortex pairs induced by impurities we argumented that
the BKT critical temperature should decrease,
justifying the MC simulations. Our results may also have
applications for granular superconducting films such as the 
ceramic high-$T_{c}$ materials. These systems could be modeled
as two-dimensional (2D) Josephson-junction arrays because such
films contain large number of Josephson boundaries between 
the small superconducting grains forming a complex Josephson-junction
network. However, the actual situation is not so ideal as the perfect array
since grains with different sizes and orientations are arranged almost 
randomly. This makes the model with vacancies more realistic than the
usual perfect array. 
The results can also be extrapolated to models with three
spin components such as 
easy-plane and XY magnets. In these cases, the problem with
impurities could be still more interesting since they have a true dynamics.
In fact, it can shed some light over
the important question about the origin of the central peak in the
dynamical spin-spin correlation function in the two dimensional anisotropic
Heisenberg model \cite{FAG,AAMB,AJE,HD,EVA1,EVA2,DG}.  
However, much work has to be done in order to
understand those effects. 
\section{Acknowledgements}
This work was partially supported by CNPq and FAPEMIG(Brazilian agencies). 
Numerical work was done at the Laborat\'orio de Computa\c{c}\~ao e 
Simula\c{c}\~ao do
Departamento de F\'{\i}sica da UFJF.
%
%------------------------------Bibliography --------------------------------
%

\end{document}